# Manufacturing, integration, and mechanical verification of SOXS

M. Aliverti*[a], L. Oggioni[a], M. Genoni[a], G. Pariani[a], O. Hershko[i], A. Brucalassi[n], G. Pignata[n,o], H. Kuncarayakti[j,k], R. Zanmar Sanchez[d], M. Munari[d], S. Campana[a], P. Schipani[c], R. Claudi[f], A. Baruffolo[f], S. Ben-Ami[i], F. Biondi[f], G. Capasso[c], R. Cosentino[h], F. D'Alessio[b], P. D'Avanzo[a], M. Landoni[a], A. Rubin[i], S. Scuderi[e], , F. Vitali[b], D. Young[t], J. Achrén[l], J. A. Araiza-Duran[m,o], I. Arcavi[p], R. Bruch[i], E. Cappellaro[f], M. Colapietro[c], M. Della Valle[c], M. De Pascale[f], R. Di Benedetto[d], S. D'Orsi[c], A. Gal-Yam[i], M. Hernandez[h], J. Kotilainen[j,k], G. Li Causi[s], S. Mattila[k], M. Rappaport[i], K. Radhakrishnan[f], E.M.A. Redaelli[a], D. Ricci[f], M. Riva[a], B. Salasnich[f], S. Smartt[t], M. Stritzinger[u], E. Ventura[h]. [a]INAF–Osservatorio Astronomico di Brera, Via Bianchi 46, I-23807, Merate, Italy; [b]INAF–Osservatorio Astronomico di Roma, Via Frascati 33, I-00078 M. Porzio Catone, Italy; [c]INAF–Osservatorio Astronomico di Capodimonte, Sal. Moiariello 16, I-80131, Naples, Italy; [d]INAF-Osservatorio Astrofisico di Catania, Via S. Sofia 78, I-95123 Catania, Italy; [e]INAF–IASF Milano Via A. Corti, 12, I-20133 Milano, fItaly; [f]INAF–Osservatorio Astronomico di Padova, Vicolo dell'Osservatorio 5, I-35122, Padua, Italy; [g]ESO, Karl Schwarzschild Strasse 2, D-85748, Garching bei München, Germany; [h]FGG-INAF, TNG, Rambla J.A. Fernandez Perez 7, E-38712 Brenã Baja (TF), Spain; [i]Weizmann Institute of Science, Herzl St 234, Rehovot, 7610001, Israel; [j]Finnish Centre for Astronomy with ESO (FINCA), FI-20014 University of Turku, Finland; [k]Tuorla Observatory, Dept. of Physics and Astronomy, FI-20014 University of Turku, Finland; [l]Incident Angle Oy, Capsiankatu 4 A 29, FI-20320 Turku, Finland; [m]Centro de Investigaciones en Optica A. C., 37150 León, Mexico; [n]Universidad Andres Bello, Avda. Republica 252, Santiago, Chile; [o]Millennium Institute of Astrophysics (MAS); [p]Tel Aviv University, Department of Astrophysics, 69978 Tel Aviv, Israel; [q]Dark Cosmology Centre, Juliane Maries Vej 30, DK-2100 Copenhagen, Denmark; [r]Aboa Space Research Oy, Tierankatu 4B, FI-20520 Turku, Finland; [s]INAF - Istituto di Astrofisica e Planetologia Spaziali, Via Fosso del Cavaliere 100, 00133, Roma, Italy; [t]Astrophysics Research Centre, Queen's University Belfast, Belfast, BT7 1NN, UK; [u]Aarhus University, Ny Munkegade 120, D-8000 Aarhus.

## ABSTRACT

SOXS (Son Of X-Shooter) is a medium resolution (~4500) wide-band (0.35 - 2.0 µm) spectrograph which passed the Final Design Review in 2018. The instrument is planned to be installed at the NTT in La Silla and it is mainly composed by five different optomechanical subsystems (Common Path, NIR spectrograph, UV-VIS spectrograph, Camera, and Calibration) and other mechanical subsystems (Interface flange, Platform, cable corotator, and cooling). It is currently in the procurement and integration phase. In this paper we present the post-FDR modifications in the mechanical design due to the various iterations with the manufacturers and the actual procurement status. The last part describes the strategy used to keep under control the mechanical interfaces between the subsystems.

*matteo.aliverti@brera.inaf.it; phone +39 347 1504637; http://www.brera.inaf.it/

## 1. INTRODUCTION

The SOXS instrument is a medium resolution spectrograph mainly developed for transients follow-up and in the procurement and integration phase. An overall view of the instrument can be found in [1], while this paper mainly focuses on the mechanical procurement and integration.

## 2. GENERAL LAYOUT

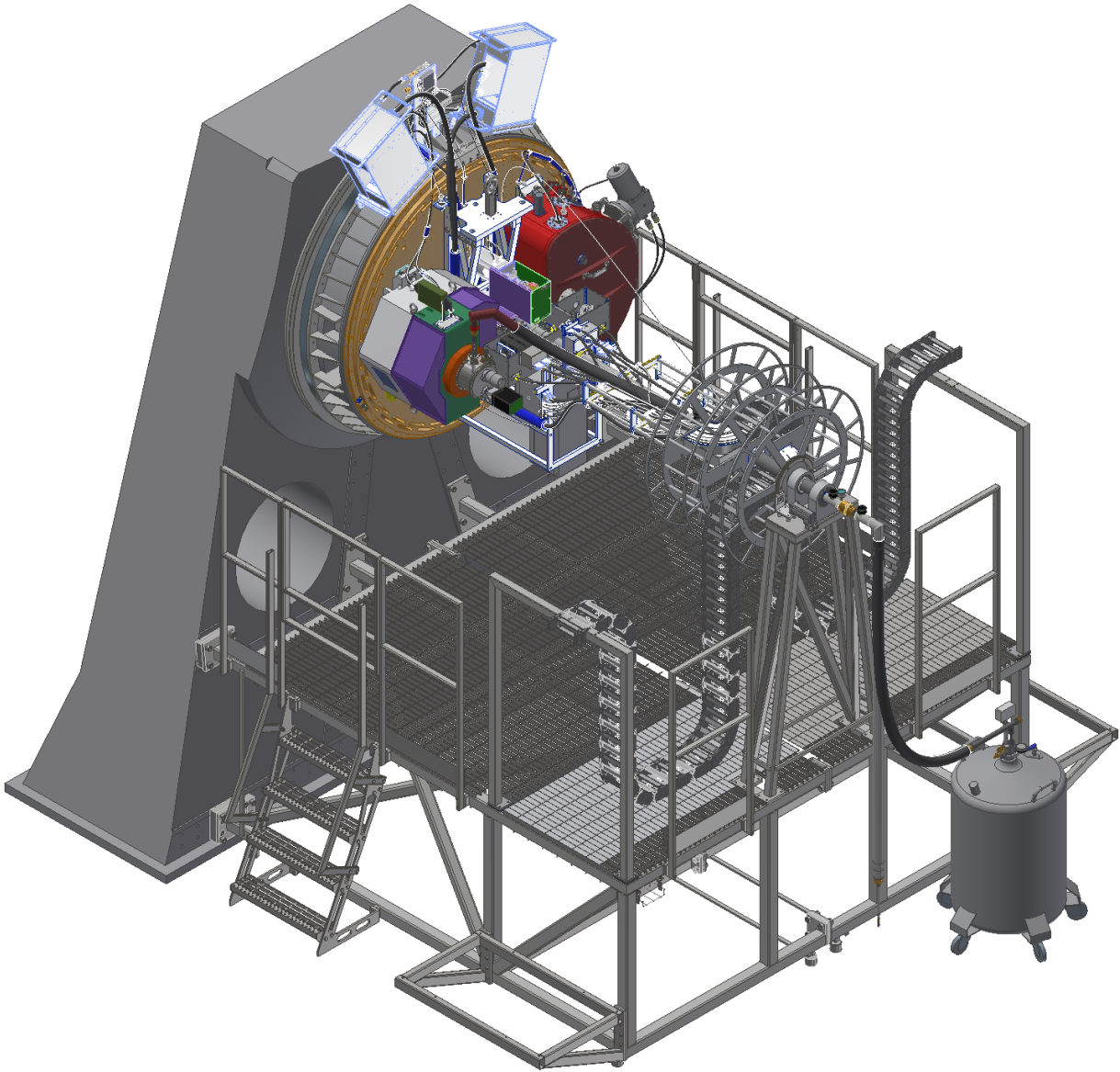

Figure 1. Overall view of the SOXS spectrograph.

The 3D model of the whole instrument can be seen in Figure 1. The rotating part of the NTT fork directly holds the interface flange (orange), the two NGCs for the UV-VIS and NIR detectors (2 light grey boxes on the top) and the cooling system holder (in between the two NGCs). The platform is connected to the non-rotating part of the NTT fork and sustain two electronic cabinets and the cable corotator. This leaves only the LN2 tank on the Nasmyth floor.

## 3. MAIN INSTRUMENT

The core of the instrument is shown in Figure 2 and is composed by the "interface flange" holding the Common Path, the UV-VIS spectrograph and the NIR spectrograph. Due to their low mass and the tight installation space, the Calibration box and the acquisition camera are connected to the Common Path structure. As the mass and momentum limitations imposed by the telescope are quite tight, and to minimize thermal misalignments in operations, most of the instrument is manufactured in aluminium 6061/6082.

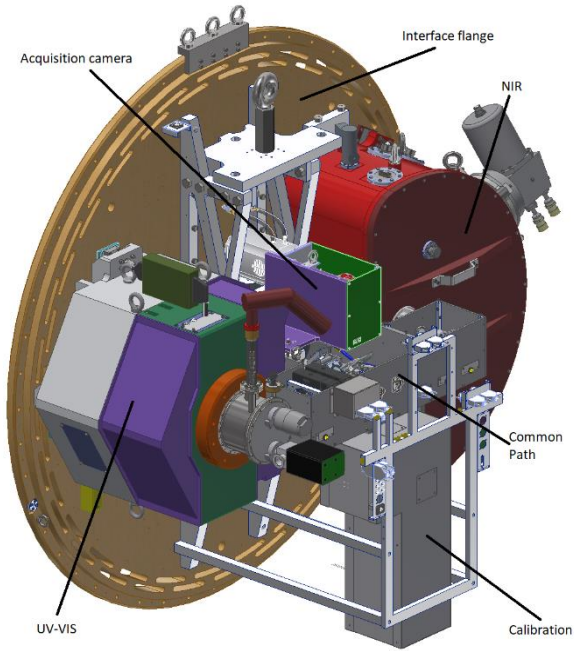

Figure 2. View of the interface flange and the 5 subsystems attached to it.

### 3.1 Interface flange

The interface flange's main role is to connect the spectrographs and the Common Path to the fork of the telescope. The element is composed by an aluminium 6082-T651 60mm thick flange with a central support and is shown in Figure 3. On the left of the figure, different features can be noticed:

- The central support, used to install the common path and to lift the whole instrument in case of maintenance using a single M30 eyebolt (plus 3 symmetric M24 eyebolts for safety reasons). The eyebolt can be regulated in the horizontal direction to co-align it with the actual centre of gravity and in vertical direction to facilitate the installation.

- Three M12 eyebolts to move the interface flange when in vertical position and without any subsystem attached.

- Two sets of supports for the installation of the UV-VIS (blue) and NIR (red) spectrographs in addition to the central plate dedicated to the Common Path.

- One cable connector support. This element defines the plane where almost all the cables are disconnected when the instrument is removed from the NTT fork

- Two couples of screws near the NIR spectrograph to facilitate its installation and to avoid sudden displacements when it is disconnected from the flange. On the UV-VIS side two rails (not shown) are installed for the same reasons while the CP does not have those elements as it is lighter and easier to move.

- Around 70 M4 holes for the routing of some cables.
- An adjustable hole (bottom left black box) and a slot (top right black box). Those elements have a diameter of 10.4mm and increase the mounting-dismounting repeatability of the instrument.

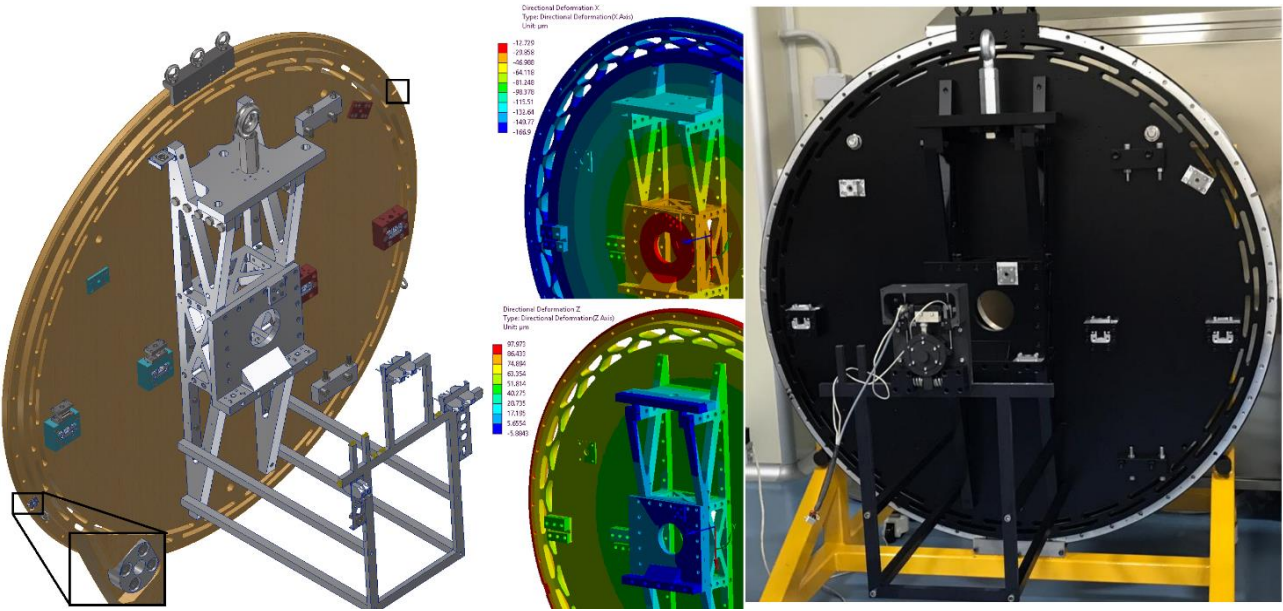

Figure 3. Left: 3D model interface flange. Centre: FEM showing the deformation of the flexures to compensate for the CTE mismatch. Right: photo of the flange after painting.

The centre of Figure 3 shows the deformations of the flexures with a temperature variation of -10°C. Those flexures strongly decrease the 'membrane effect' due to the different CTE of the instrument (aluminium) and the telescope (structural steel) and remove the flatness errors on the interface surfaces. A more detailed analysis can be found in [2].

On the right of Figure 3 a picture of the interface flange with all the parts except for the interface surfaces painted with a high strength paint.

### 3.2 Common Path (CP)

The overall view of the CP is shown in Figure 4. Following the light coming from the telescope, we have a shutter, the calibration stage, the camera stage and the dichroic. The reflected light goes to the UV-VIS spectrograph through a mirror, the ADC, a tip-tilt mirror and the UV-VIS pupil stop. The refracted light, instead, goes to the NIR spectrograph through a mirror, a tip-tilt mirror and a refocuser. Both the outputs have two baffle to avoid spurious light from entering the spectrographs.

Other important elements are the kinematic mounts (in blue) used to install the CP on the interface flange and the Acquisition Camera on the CP. On the bottom part of the CP structure, another set of kinematic mounts identical to the acquisition camera one is used to install the calibration unit.

The Thorlabs pinholes supports used for the pinhole alignment are indicated in green (a fifth one is behind the shutter) as for the Thorlabs magnetic supports used to mount a camera on the exit focal planes.

In the same figure, a picture of the completely installed CP without the optics to test the installation of all the components is shown together with the pre-aligned one. The first picture has been taken after the painting with Z306. The latter has all the optical elements installed and the pinholes and the dichroic aligned. The ADC has been removed and shipped to the optical manufacturer for the installation and pre-alignment of the optics. A dummy plate has been installed on the acquisition camera selector, as the optical element was not yet ready during the CP integration.

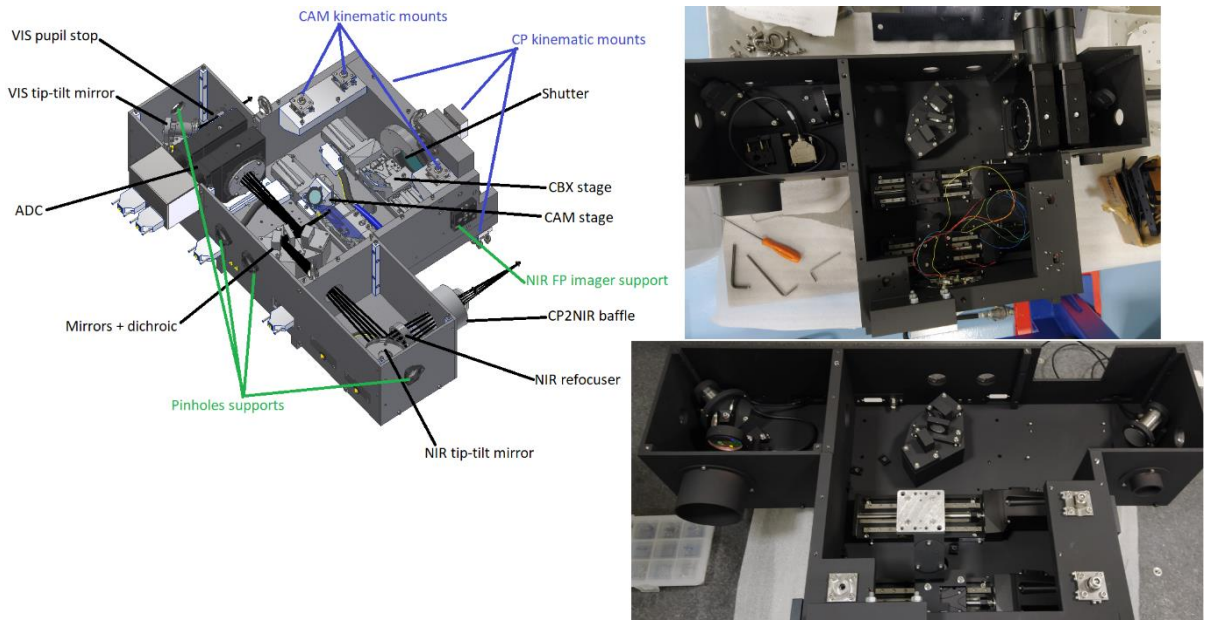

Figure 4. Overall view of the Common Path. Left: 3D model with the main components highlighted. Top right: picture with all the components installed. Bottom right: picture taken before the shipment to Padua for the final alignment.

The first element is a mono-stable normally closed Vincent Associates Uniblitz CS65E1T1 shutter. Preliminary tests showed that the actuator can heat up if kept open for a long time (see Figure 5). Therefore, a black anodized protective cover has been installed in front of it in order to mask the hot spots and to dissipate the heat.

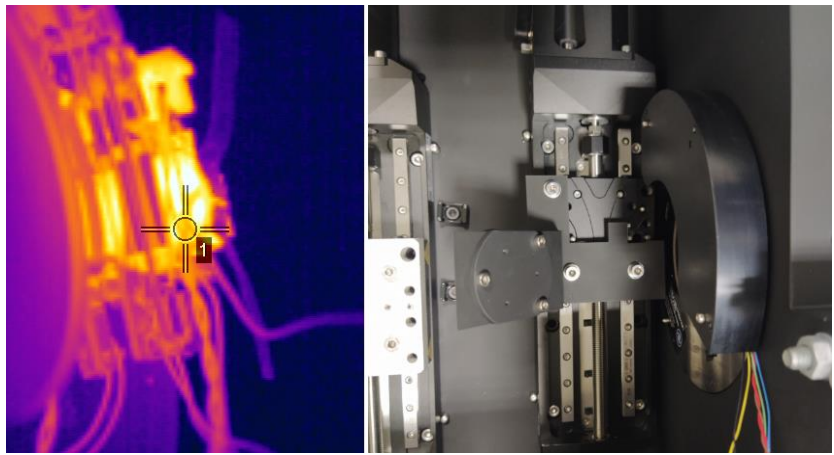

Figure 5. Left: thermal measure of the shutter. Right: shutter installed inside the CP with its cover.

The stages inside the CP are two PI L-406.40DD10 (camera selector and calibration selector), two Owis DMT 100-D53-HiDS (ADC), two PI S-330.2SL (tip-tilt stages) and one PI M-111.1DG1 (NIR refocuser). The static and dynamic characterization of those elements using a CMM and a Laser Tracker was foreseen but, due to the COVID-19 pandemic, those tests have been prepared (see Figure 6) but not performed.

The first stage encountered is the calibration selector (see Figure 5), which has a 135° support holding a cylindrical mount with the Edmund mirror glued. The second element is the camera selector and is shown in Figure 7. The support has been manufactured and will soon be glued to the fused holed mirror. Said mirror has two 45° holes and two laser-cut aluminium apertures (one quasi-elliptical and ne quasi-rectangular) glued on it by the optical manufacturer. The element will be glued to the support using 6 spots of RTV glue after the mechanical alignment of the mirror tip, tilt and clock using 6 screws and the CMM.

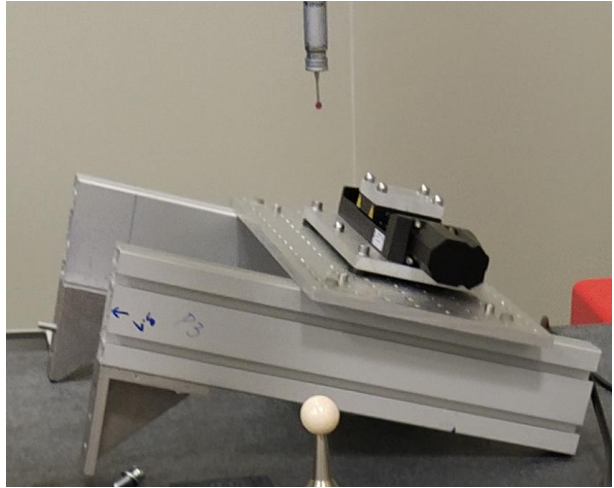

Figure 6. Linear stage ready for the tests under the CMM.

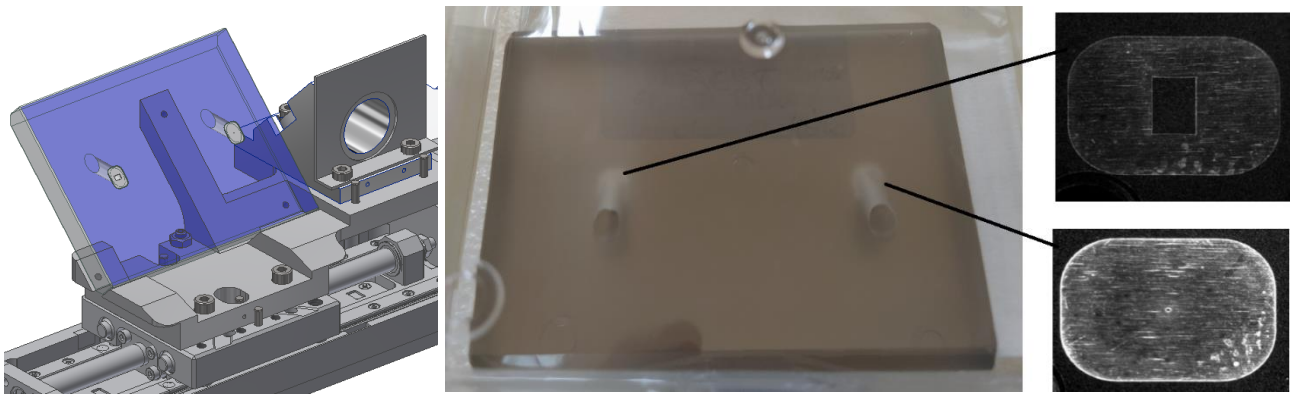

Figure 7. Camera selector mirror. Left: mounted on the camera selector stage. Centre: view of the back surface. Right: pictures of the two holes mounted on the front face.

On the UV-VIS side, the two rotary motors part of the ADC have been sent to the optical manufacturer. Before that, they have been pre-aligned with an AACMM (see Figure 8) to allow for the installation of the optical elements. A finer hybrid optical and mechanical alignment will be performed once the optical elements will be delivered.

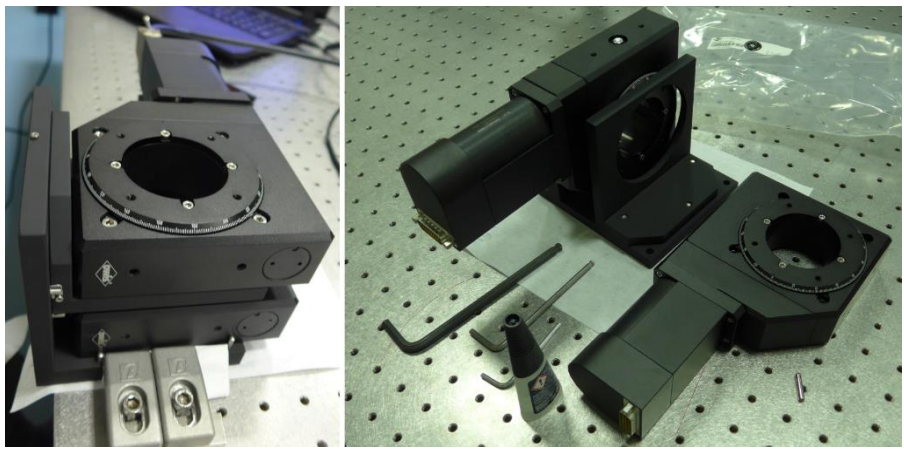

Figure 8. View of the two ADC motors mounted on their supports.

All the other CP elements have been glued following different strategies (see Figure 9):

- the NIR refocuser has been glued with 3 radial spots of 3M 2216
- the dichroic with 6 angular spots of RTV adhesive
- the small mirrors with 3 RTV spots on the back
- the large mirror with 6 RTV spots on the back

The adhesive thicknesses have been set with calibrated shims and different tests have been performed to ensure the strength of the glues, the adhesion to the substrate, the final spot thickness and diameter and the absence of bubbles.

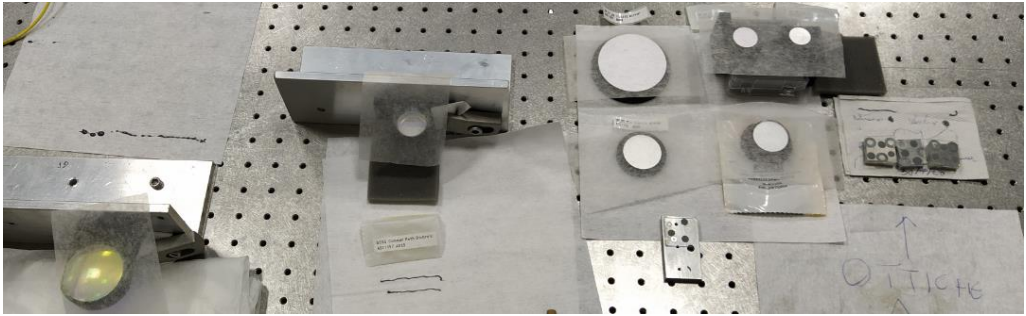

Figure 9. All the optical elements during the adhesive curing. From left to right: NIR refocuser, dichroic, tip-tilt mirrors, CBX+NIR+UVVIS mirrors, test pieces.

After the curing, the elements have been tested under the CMM (see Figure 10), installed inside the CP (as discussed in section 5) and the cables routed. As a final test, the CP has been installed and uninstalled from the interface flange a number of times.

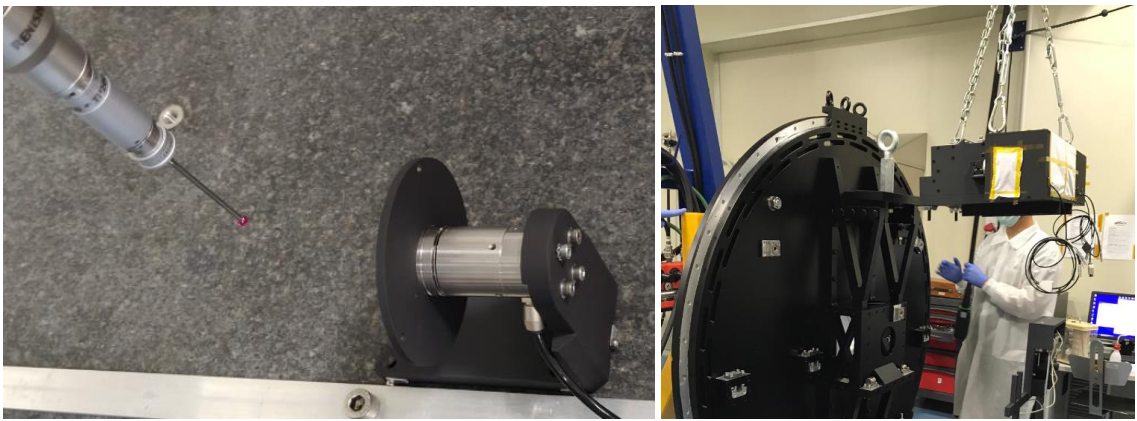

Figure 10. Left: NIR tip-tilt mirror during the CMM verification process. Right: installation test on the interface flange.

### 3.3 Calibration Unit (CBX)

The calibration unit is composed by a 'light box' and an 'optic box'. The optic box is directly connected to the CP via three kinematic mounts, which can be seen in Figure 11 (left). More details on this subsystem can be found in [3].

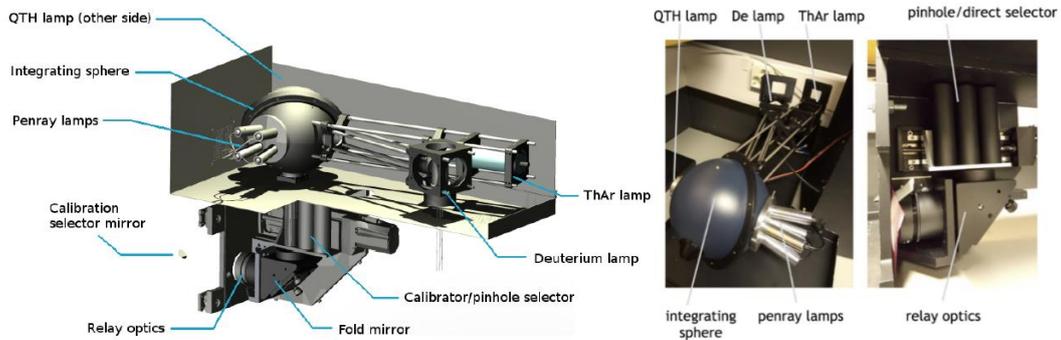
Figure 11. CBX. 3D view of the calibration box (left), light box (centre) and optic box (right).

### 3.4 Acquisition Camera (CAM)

The acquisition camera unit is shown in Figure 12. As in the CBX case, also this subsystem is connected to the CP via three kinematic mounts, which can be seen in the bottom of the figure. More details can be found in [4].

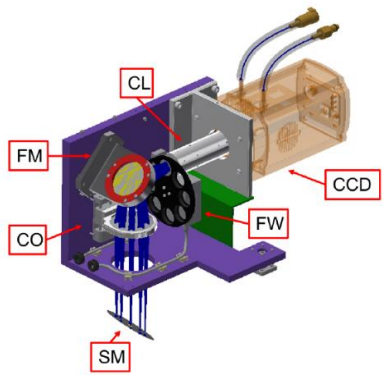
Figure 12. 3D view of the acquisition camera.

### 3.5 UV-VIS spectrograph

The UV-VIS spectrograph, shown in Figure 13, is directly connected to the interface flange via three large kinematic mounts. More details on this subsystem can be found in [5] and [6]. Another notable interface is the 90° Johnston coupling (dark red in Figure 13) used to transfer the liquid nitrogen from a tank to the CFC via a rotating LN2 line described in section 4.2.

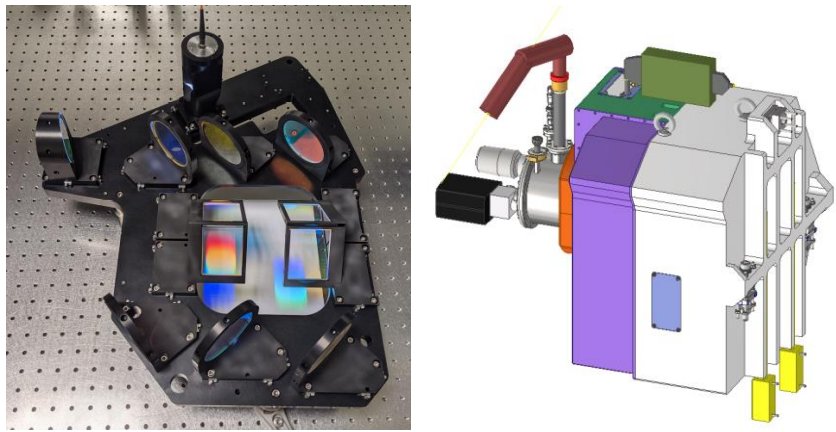
Figure 13. Left: top view of the feed part of the UV-VIS spectrograph. Right: overall view of the UV-VIS spectrograph.

## 3.6 NIR spectrograph

The NIR spectrograph is currently in the manufacturing and integration phase. The Vacuum Vessel (Figure 14) has been manufactured and painted and it is kept under vacuum until all the elements are ready to be installed. In the same figure, a 3D view of the subsystem is shown with the location of the flexures and the 3 main sub-benches:

- The 'slit sub-bench' includes the slit, the pupil stop, the first lens and the first mirror,
- The 'dispersing sub-bench' includes the collimator lens, the 3 prisms and the grating,
- The 'camera sub-bench' includes the 3 camera lenses and the filter.

The only optomechanical components not located on those sub-benches are the two spherical mirror and the detector support.

Almost all the components in the subsystem are made of aluminium 6082 with the notable exceptions of the insulators (G10-FR4), the straps (Copper), the flexures and the slit motor (Titanium alloy).

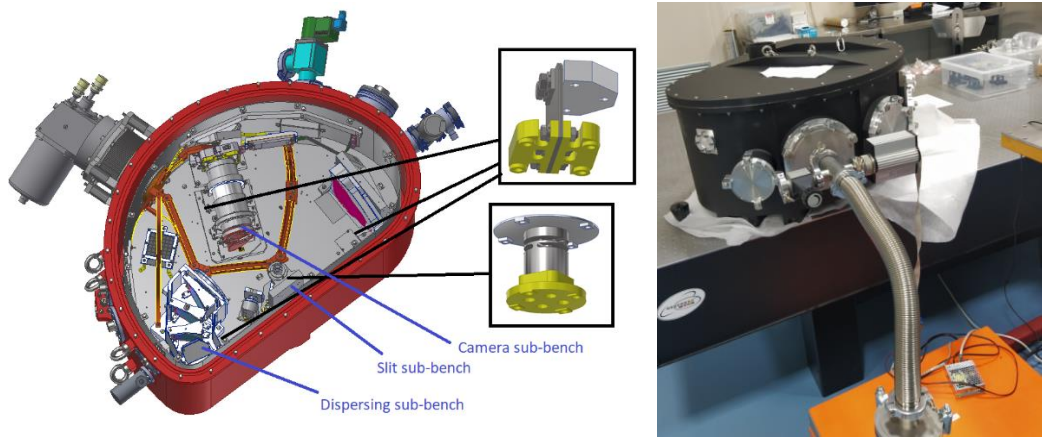

Figure 14. Left: overall view of the NIR spectrograph with the main sub-bench and the flexures location highlighted. Right: vacuum vessel during the vacuum tests.

All the mechanical elements have been received with the exception of the dispersing sub-bench ones and are shown in Figure 15. Due to some modification on the collimator lens requested by the manufacturer the dispersing sub-bench has been completely redesigned and, therefore, the mechanical manufacturing has started later. All the elements are mounted on a dummy bench used for training and for the first in-air alignment.

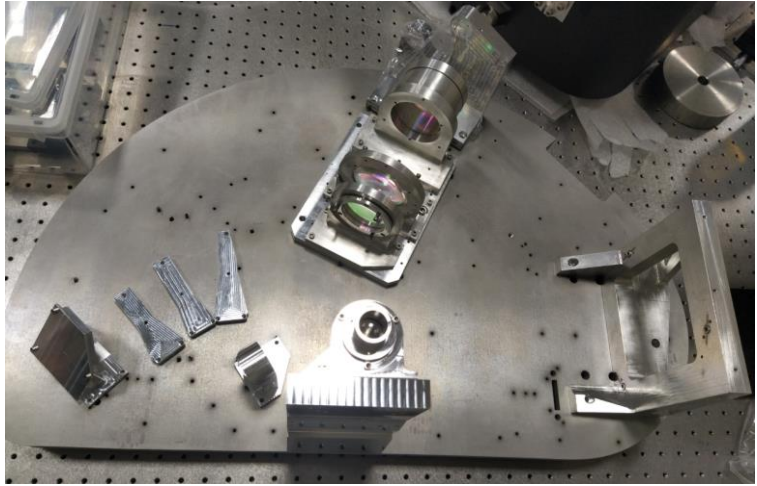

Figure 15. Manufactured optomechanical elements placed on the dummy bench.

The slit sub-bench can be seen in Figure 15 and Figure 16. This subsystem includes, from top to bottom, the baffle, the pupil stop, a small lens kept in position with 3 springs, the slit motor, another baffle and the first flat mirror. Different design solutions have been evaluated for the slits: the etching technology has been discarded for different reasons including economic ones while the original plan to use laser cut on titanium foil has failed due to the poor thermal conductivity of the material. After a long series of tests, the best results have been obtained with a 0.2mm aluminium foil and a femtosecond pulsed laser.

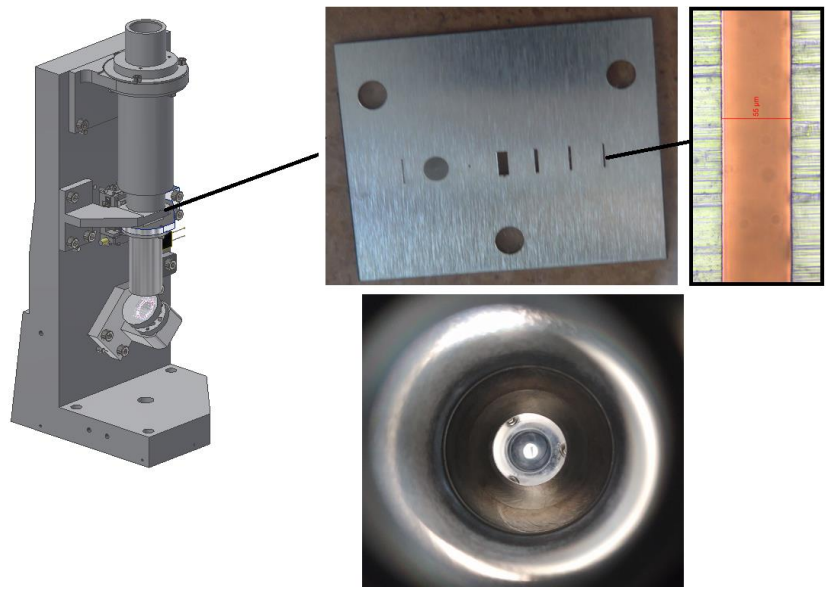

Figure 16. Slit sub-bench. Left: 3D view of the element. Top right: view of the slits. Bottom right: view from the top; the small aperture at the end is the aperture on the first lens support.

The dispersing sub-bench (Figure 17) is currently under manufacturing but a 3D printed version has been produced together with dummy aluminium prisms in order to start the training for the alignment operations. Those dummy prisms will be also used to train on the final system and to perform the mechanical alignment of the dispersing sub-bench without using the actual prisms.

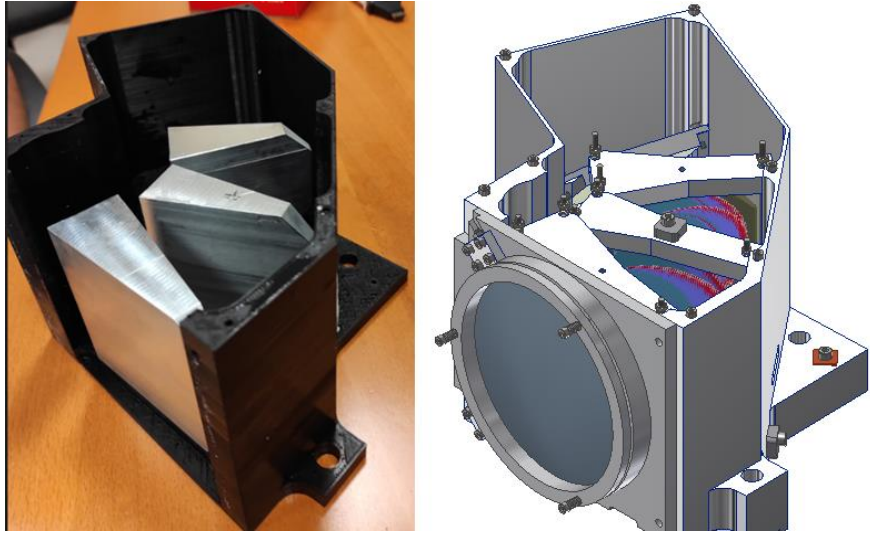

Figure 17. Dispersing sub-bench. Left: 3D printed bench with dummy prisms. Right: 3D view of the dispersing sub-bench.

The camera sub-bench (Figure 18) is ready for the painting process. Before this operation the optomechanical elements will be tested with different hot (393K) – cold (150K) cycles using a dedicated vacuum vessel and some dummy lenses to prove the repeatability and the safety of the conical springs system already presented in [7].

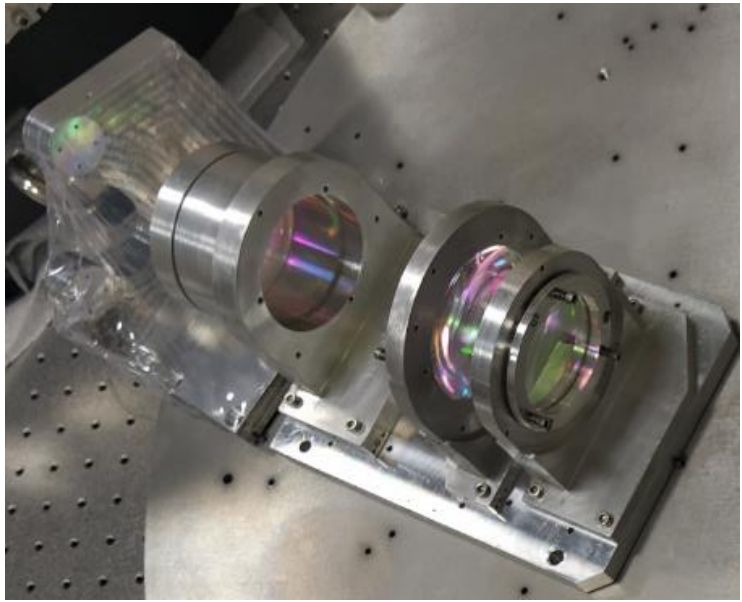

Figure 18. Camera sub-bench with optical elements inside and detector support on the back.

The aluminium mirrors supports are already in house and the mirrors are in the final stages of manufacturing. The small ones (Figure 19, right) include a wineglass foot and 3 slots. Those features reduce the deformations induced on the optical surface by the mounting and allow for looser tolerances on the manufacturing of the mirror mounts. For the large collimator mirror (Figure 19, left) the wineglass foot option was not feasible and, therefore, 3 flexures have been manufactured on the mechanical support (see Figure 15). Those 3 flexures are different from each other to counter-rotate the mirror with respect to the other optical elements; this should keep the image stable on the detector when the instrument rotates.

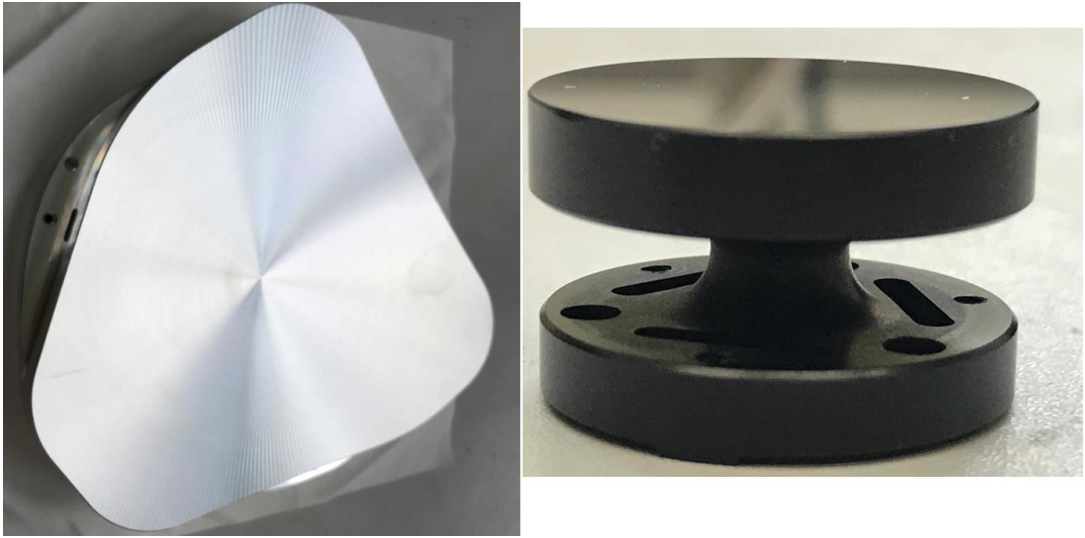

Figure 19. Aluminium mirrors under manufacturing.

Finally yet importantly, the NIR telescope simulator, shown in Figure 20, has been manufactured and aligned. It is composed by a fibre support, two parabolic aluminium mirrors and a pupil stop. The element will be mounted and aligned on the NIR vacuum vessel cover during the spectrograph alignment.

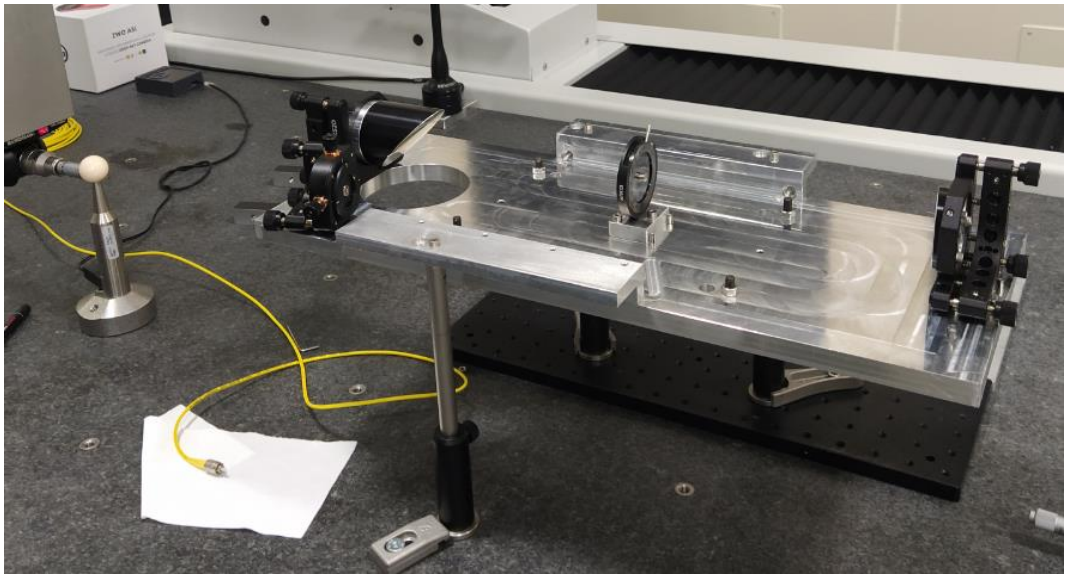

Figure 20. Telescope simulator system during the alignment.

## 4. TOOLS AND AUXILIARY ELEMENTS

In addition to the core of the instrument, other tools and auxiliary elements have to be manufactured. A brief list is shown hereafter.

### 4.1 Platform

The 3D design of the platform is shown in Figure 21. It is divided in a number of parts including two mirrored main bodies, a corotator support, two cable supports, two rack supports, 6 guardrails, 2 step ladders (each with 2 guardrails) and two doors. The main modifications done after the FDR are related to the manufacturability of the pieces, as the milling will be performed on the whole platform, and to the safety with the replacement of the stairs with stepladders and the addition of toe-plates. Other minor modifications have been applied to comply with the relevant international standards.

In-between the two main pieces some regulation plates are foreseen. The 6 interfaces with the NTT fork have an intermediate plate for the final regulation once the system will reach the telescope. Three vertically tuneable and removable ball transfer units are foreseen for below each of the two main parts to allow for small movements during the integration.

The platform is still under procurement phase and it is expected to be delivered in 2021.

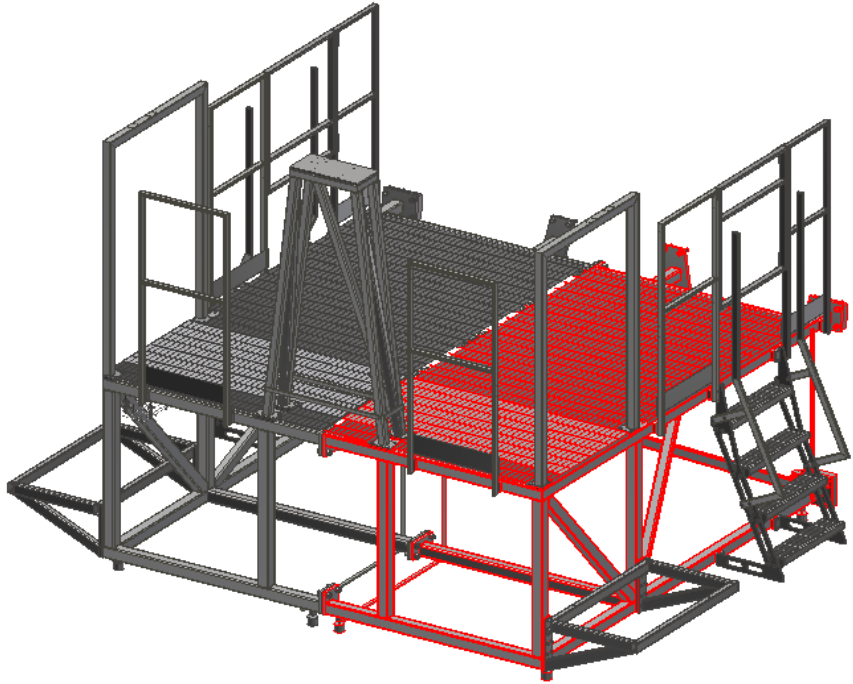

Figure 21. Isometric view of the platform. In red one of the main components.

**4.2 Corotator, corotator feedback and LN2 line**

The corotator and the corotator feedback are shown in Figure 22. The corotator will be installed on the corotator support visible in Figure 21 using two spherical joints. The corotator is composed by a number of black anodized aluminium parts held together by a central hollow steel tube, which is connected to the spherical joints. The transmission of the motion is done mounting a 120-tooth external ring gear on the last plate of the corotator and a 26-tooth pinion together with a 60:1 epicyclical reducer on the Schneider electric BSH0551T11A2A motor.

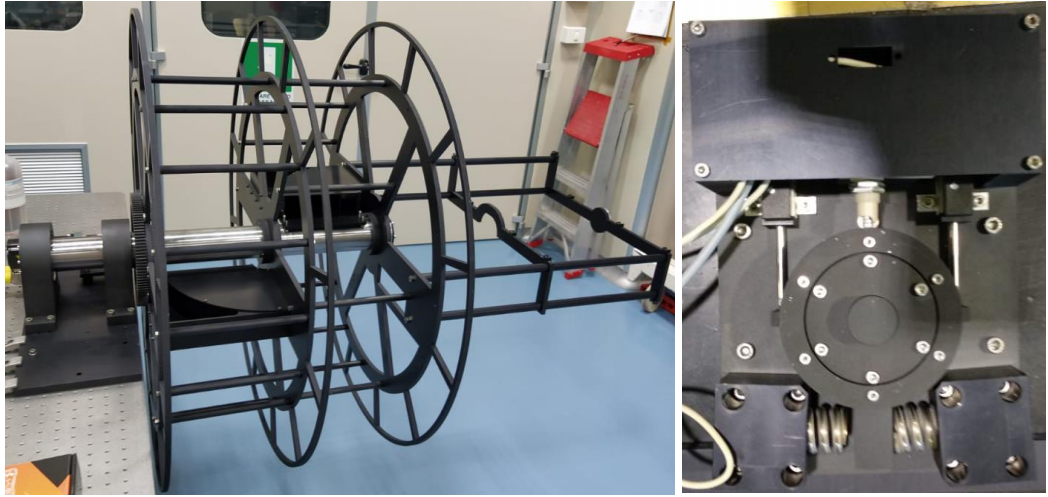

Figure 22. Left: overall view of the corotator. Right: front view of the feedback system.

The feedback part is the one that checks the relative rotation between the instrument and the cable corotator. It is mounted on the interface flange cables support (see Figure 3) and it is connected to the corotator using two bellows. A short shaft is supported by a double row angular contact ball bearing mounted on the fixed support.

The synchronization between corotator and Nasmyth has three levels:

1. Two linear potentiometers (Tekkal TR/25) will give the feedback to the motor that will control the speed of the corotator to match the Nasmyth one.
2. One Omron switch ZC Q2255 that cut off the power from the corotator and the Nasmyth motor in case the misalignment between the 2 exceed the ±10° (tuneable value in amplitude and zero location)
3. A crown gear will mechanically fix the rotation between the two in case the misalignment exceed the ±16° value.

Two springs are activated for synchronization errors >±3° to avoid any unwanted switch activation when the corotator motor is turned off and to damp the engagement forces between instrument and corotator in case of failure.

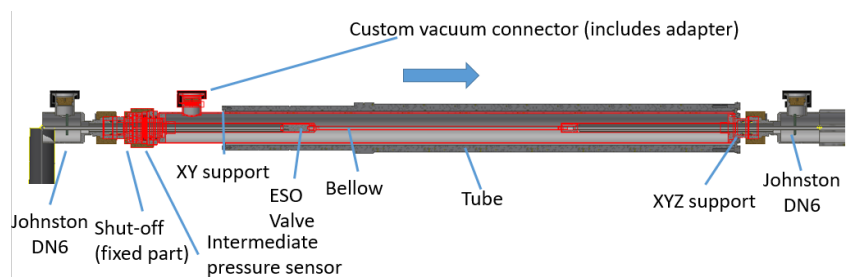

Figure 23. Section of the rotating part of the LN2 line.

The liquid nitrogen line, shown in Figure 23, has been designed by a private company and it is in the manufacturing phase. It will be composed by 3 parts: 2 flexible tubes with male Johnston couplings and one rotating part installed inside the hollow tube of the corotator. The peculiar part of this system is that the rotation motion and the vacuum sealing is done by a series of O-rings instead of a ferrofluidic feedthrough. To keep under control the O-ring wearing, a pressure sensor has been installed after the first one.

### 4.3 Interface flange support

The interface flange support is shown in Figure 24 and is a structural steel frame with 4 wheels. The flange is placed on a central cylindrical rubber support (bottom centre) and the clock is controlled by two rubber cylinders. Once it is in position it is secured by 3 M24 bolts.

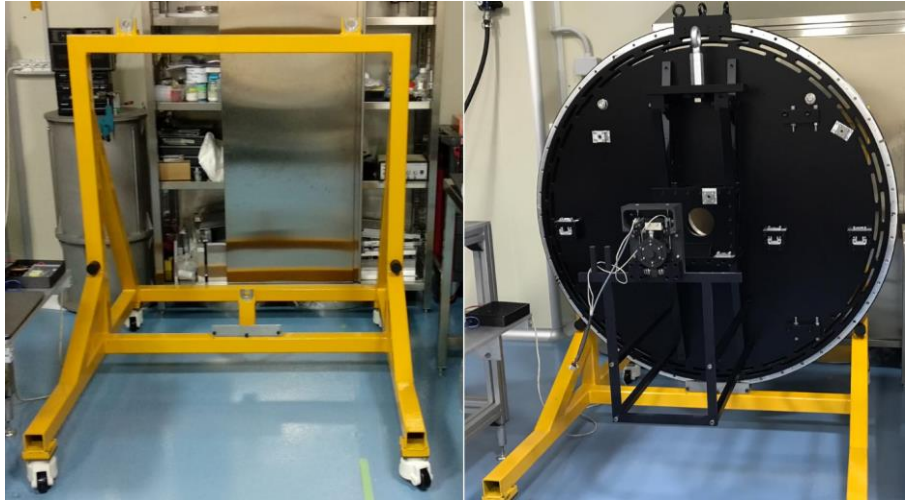

Figure 24. Interface flange support with and without the plate installed.

### 4.4 NIR-CP-UVVIS support

The rotating platform shown in Figure 25 has two possible positions: a horizontal one and the vertical one. The horizontal position will be used for the integration or the maintenance of the spectrographs while moving in the vertical position allows for the installation of the spectrographs on the interface flange.

This platform will also be used to verify the NIR spectrograph performances in two different gravity conditions.

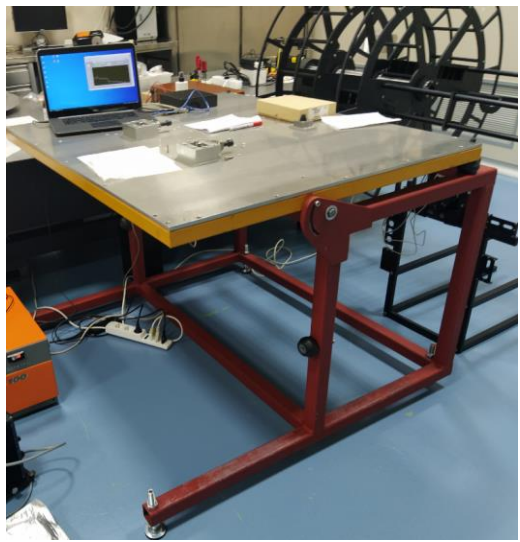

Figure 25. Rotating platform in the 'horizontal position'.

## 5. INTEGRATION

A large number of alignment and verification operation have been postponed due to the pandemic related lock-down. The work, then, has been focused on the Common Path and the interface flange pre-alignment.

### 5.1 CP preliminary alignment

The CP pre-alignment consisted in 4 main parts: the alignment of the kinematic mounts, the alignment of the pinholes supports, the alignment of the of the dichroic and the pre-alignment of the optomechanical elements.

The kinematic mounts have been aligned using the CMM and using as a reference the 3 precisely machined surfaces highlighted in Figure 26.

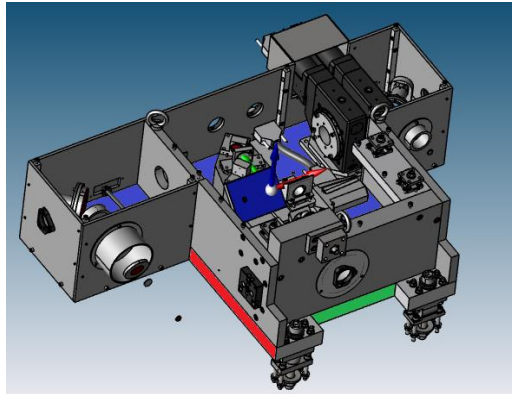

Figure 26. CP and the three planes used to define the SOXS reference system.

The only exception are the kinematic mounts holding the CBX. They have been measured with an AACMM due to their low accessibility. The results after the alignment are shown in Table 1 and have been considered satisfactory for different reasons: deviation on the 'Main' elements can be easily compensated on the interface flange, the CAM does not have tight tolerances and the CBX will be aligned directly on the CP.

Table 1. Measured deviation of the CP Kinematic Mounts.

|  | FLA/CYL/SPH [mm] | DIA [mm] | X [mm] deviation | Y [mm] deviation | Z [mm] deviation |
|---|---|---|---|---|---|
| Sphere Main | 0.004 | 30+0.004 | +0.058 | -0.082 | +0.009 |
| Cylinder Main | 0.015 | 30+0.017 |  | -0.051 | -0.014 |
| Plane Main | N/A |  |  | -0.075 |  |
| Sphere CAM | 0.002 | 15+0.010 | +0.091 | -0.013 | -0.006 |
| Cylinder CAM | 0.003 | 15-0.005 | +0.116 |  | +0.024 |
| Plane CAM | N/A |  |  |  | -0.013 |
| Sphere CBX (AACMM) | 0.015 | 15+0.000 | -0.014 | -0.062 | +0.022 |
| Cylinder CBX (AACMM) | 0.032 | 15+0.000 | +0.061 |  | +0.017 |
| Plane CBX (AACMM) | 0.012 |  |  |  | -0.018 |

In order to perform the optical alignment of the CP in Padua (presented in [8]) six accurately placed pinholes were needed. The chosen strategy consisted in the characterization of the position of each pinhole inside its magnetic support with a microscope (operation performed in Padua) and the alignment and gluing of the other half of the magnetic support in Milan. The obtained deviation with respect to the desired values are shown in Table 2.

Table 2. Measured nominal values and deviations of the pinholes.

|  | Position X [mm] | Position Y [mm] | Position Z [mm] |
|---|---|---|---|
| Pinhole 1 (in axis with entrance) | +91.425(-0.012) |  | -21.021(-0.007) |
| Pinhole 2 (off axis with entrance) | +16.437(-0.009) |  | -21.290(+0.005) |
| Pinhole 3 (entrance) | -16.389(+0.010) |  | -21.370(-0.011) |
| Pinhole 4 (NIR) |  | +121.794(-0.004) | -21.058(+0.000) |
| Pinhole 5 (VIS) |  | +51.402(+0.005) | -21.302(+0.006) |

The alignment of the dichroic has been done measuring the surfaces of the optical element with the CMM and shimming the optomechanical mount. This operation has been performed in about one hour and resulted in a focus error of 0.013mm, a tip-tilt of 0.871-0.115mrad (first surface, high reflection in the visible) and 1.285-0.117mrad (second surface, anti-reflection in the infrared). At the end of the operation, six reference points have been acquired on the optomechanical mounts for potential future dismantling of the element.

The last operation consisted in the measurement of all the glued optomechanical elements to ensure that they could be placed inside the CP structure in a range of ±0.2mm and ±3mrad. All the elements satisfied those requirements with deviations typically below ±0.1mm and ±1mrad. On the tip-tilt mounts and the CBX selector mirror the deviations have been further reduced varying the internal shimming.

## 5.2 Interface flange alignment

A first round of alignment of the interface flange kinematic mounts has been performed with an AACMM (see Figure 27).

The results can be seen in Table 3 and have been obtained with careful shimming of all the 9 kinematic elements trying to optimize the overall clock of the flange, the relative position of the 3 sets and to minimize tip-tilt and clock of each subsystem.

The UV-VIS spectrograph set has been placed considering its nominal position at 20°C (typical integration temperature).

The same environmental conditions have been considered for the CP set but the positions have been modified to compensate the kinematic mount deviations shown in the section 5.1.

The NIR spectrograph set has been placed, again, in its nominal position but the deformations due to the vacuum and the cold temperatures inside the vacuum vessel (obtained by thermoelastic FEA) have been taken into account.

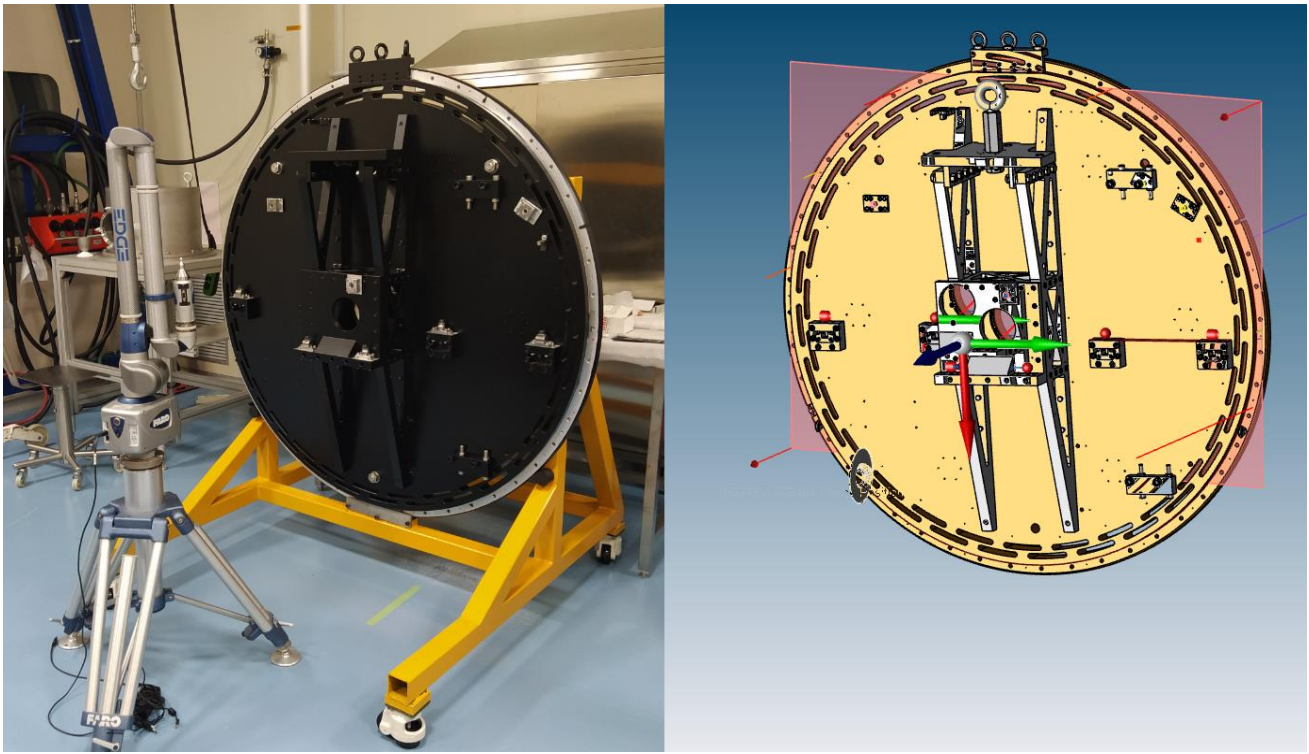

Figure 27. Left: interface flange during the alignment process. Right: the features measured in red.

Table 3. First alignment results of the interface flange kinematic mounts.

|  | Deviation X [mm] | Deviation Y [mm] | Position Z [mm] | Notes |
|---|---|---|---|---|
| CP Sphere | -0.001 | -0.000 | +0.000 | Actual, 293K, 1bar |
| CP Cylinder | -0.001 |  | -0.021 | Actual, 293K, 1bar |
| CP Plane |  |  | -0.008 | Actual, 293K, 1bar |
| NIR Sphere | -0.005 | -0.005 | -0.037 | Nominal, 150K, 0bar |
| NIR Cylinder | -0.029 | | +0.030 | Nominal, 150K, 0bar |
| NIR Plane | | | +0.038 | Nominal, 150K, 0bar |
| VIS Sphere | +0.001 | -0.011 | +0.054 | Nominal, 293K, 1bar |
| VIS Cylinder | +0.039 | | -0.016 | Nominal, 293K, 1bar |
| VIS Plane | | | +0.050 | Nominal, 293K, 1bar |

This is considered the first iteration as the position of the kinematic mounts will be corrected once the actual distance between the spectrographs focal planes and their respective kinematic mounts will be measured in the operative conditions (293 or 150 K) together with the actual positions of the common path exit focal planes.

## 6. CONCLUSIONS

The integration of SOXS is proceeding despite the challenging posed by the COVID-19 pandemic. From the mechanical point of view, the procurement is almost completed and the alignment of the NIR spectrograph, which is on the critical path, should start at the beginning of 2021. A summary of the operation performed on the CP for the alignment have been presented along with the results. The results obtained on the alignment of the interface flange kinematic mount have been also presented. The next steps to be performed in Milan will include the mechanical alignment and optical verification of the NIR spectrograph while all the tools and subsystem will be shipped to Padua for the final integration.